\documentclass{article}
\usepackage{ijcai22}

\usepackage{times}
\usepackage{soul}
\usepackage{url}
\usepackage[hidelinks]{hyperref}
\usepackage[utf8]{inputenc}
\usepackage[small]{caption}
\usepackage{graphicx}
\graphicspath{ {./images/} }
\usepackage{CJKutf8}
\usepackage{multirow}
\usepackage{amsmath}
\usepackage{amsthm}
\usepackage{booktabs}
\usepackage{algorithm}
\usepackage{algorithmic}
\usepackage{xcolor}
\usepackage[switch]{lineno}
\usepackage{fontawesome}

\title{Interpretable Tsetlin Machine-based Premature Ventricular \\ Contraction Identification }

\author{Jinbao Zhang$^1$\and 
Xuan Zhang$^2$\and
Lei Jiao$^3$\and
Ole-Christoffer Granmo$^3$\and\\
Yongjun Qian$^4$\and
Fan Pan$^1$\textsuperscript{\faEnvelopeO} \affiliations
$^1$College of Electronics and Information Engineering, Sichuan University, Chengdu, China\\
$^2$Norwegian Research Center (NORCE) AS, Grimstad, Norway\\
$^3$Center for Artificial Intelligence Research, University of Agder, Grimstad, Norway\\
$^4$Department of Cardiovascular Surgery, West China Hospital, Chengdu, China
\emails
panfan@scu.edu.cn
}
\begin{document}
\begin{CJK}{UTF8}{gbsn}
\maketitle

\begin{abstract}
Neural network-based models have found wide use in automatic long-term electrocardiogram (ECG) analysis. However, such black box models are inadequate for analyzing physiological signals where credibility and interpretability are crucial. Indeed, how to make ECG analysis transparent is still an open problem. In this study, we develop a Tsetlin machine (TM) based architecture for premature ventricular contraction (PVC) identification by analyzing long-term ECG signals. The architecture is transparent by describing patterns directly with logical AND rules. To validate the accuracy of our approach, we compare the TM performance with those of convolutional neural networks (CNNs). Our numerical results demonstrate that TM provides comparable performance with CNNs on the MIT-BIH database. To validate interpretability, we provide explanatory diagrams that show how TM makes the PVC identification from confirming and invalidating patterns. We argue that these are compatible with medical knowledge so that they can be readily understood and verified by a medical doctor. Accordingly, we believe this study paves the way for machine learning (ML) for ECG analysis in clinical practice.

 %Premature ventricular contractions (PVC) can increase the risk of developing irregular heart rhythms (arrhythmias) or weakening of the heart muscle (cardiomyopathy). It may even lead to chaotic, dangerous heart rhythms and possibly sudden cardiac death. Automatic long-term electrocardiogram (ECG) analysis algorithm is a hot spot in recent years, and neural network models are widely used in this. However, in addition to being accurate, the discrimination of physiological signals should have reasonable criteria to increase the credibility of the model. It is difficult for black box models such as neural networks to give reasonable criteria. So we expect the Tsetlin machine~(TM) to give reasonable criteria based on its strong mathematical interpretation when performing classification. In this study, TM was used for ECG signal discrimination for the first time. In the experiment, a convolutional neural network~(CNN) is used to compare with TM in addition to giving a reasonable explanation of the discrimination. In the end, both TM and the CNN achieved an accuracy rate of more than 93\% in the MIT-BIH database. Moreover, we visualized the basis for TM to make judgments, and made corresponding explanatory diagrams that can help people review and understand.

\end{abstract}

\section{Introduction}

%Cardiovascular diseases (CVDs) are the foremost cause of human death worldwide, which can lead to over 31\% of deaths every year. With the progressive aging of populations worldwide, the number of patients with CVDs may continue to increase. It is estimated that the number of deaths due to CVDs will increase from 17 million in 2016 to 24 million in 2030~\cite{1}. For this reason, monitoring and preventing CVDs in advance have become one of the important tasks for many countries~\cite{2}.
Cardiovascular diseases~(CVDs) are the leading cause of mortality globally, encompassing more than 32\% of deaths annually. Arrhythmia is one of the major causes of CVDs, associated with abnormal initiation or propagation of a wave of cardiac excitation. As a common type of arrhythmia, premature ventricular contraction~(PVC) is caused by irregular contractions that start in the right or left ventricle instead of the atria~\cite{3}. Frequently experiencing PVC might increase the risk of developing cardiomyopathy or weakening of the heart muscle. If accompanied by heart disease, frequent PVC can even lead to chaotic and dangerous heart rhythms, and possibly sudden cardiac death~\cite{4}.

In order to diagnose PVC, the patient needs to do an electrocardiogram~(ECG) test or wear a Holter monitor to record the ECG signal. Figure 1 shows a typical ECG signal. Clinicians can detect PVC from the signal by identifying an abnormal and extensive QRS complex that occurs earlier than expected in the cardiac cycle. Such a diagnosis must be precise and reasonably explained to be sufficiently credible. Accordingly, the procedure imposes high workload on the clinicians. For this reason, it is clinically crucial to develop an automatic, accurate, and explainable PVC identification method.

%Arrhythmia is a common CVDs, which refers to a series of rhythm and/or waveform irregular. As one of the most common arrhythmias, premature ventricular contraction~(PVC) is caused by premature ectopic beats in the right or left ventricle~\cite{3}. Frequent PVC and multisource PVC detection have important clinical significance~\cite{4}. Clinicians typically detect and differentiate PVCs by observing rhythmic changes and subtle morphological changes in the electrocardiogram~(ECG) signal. Such a judgment requires an accurate result, and a reasonable explanation to increase the credibility of the judgment.

Machine learning~(ML) has been widely applied to the medical field with impressive outcomes, including diagnosis of pneumonia~\cite{5}, assessment of CVDs risk~\cite{6}, detection of acute intracranial haemorrhage~(ICH)~\cite{7}, and blood pressure measurement~\cite{8}. Recently, ML has also been employed to diagnose arrhythmia from ECG signals~\cite{9,10,11,12}. The state-of-the-art for detecting and classifying arrhythmia now uses deep neural networks~(DNNs)~\cite{13} that achieve a precision of 83.7\%, which exceeds the performance of an average cardiologists (78.0\%).  Similarly, a multi-label feature selection method with an accompanying ML model was introduced for classification of arrhythmia in ~\cite{14}, providing an average precision of 84.6\%. 

\begin{figure}[b]
\begin{center}
    \setlength{\abovecaptionskip}{-0cm} %调整图片标题与图距离
    \setlength{\belowcaptionskip}{-0.1cm} %调整图片标题与下文距离
\includegraphics[scale= 0.6]{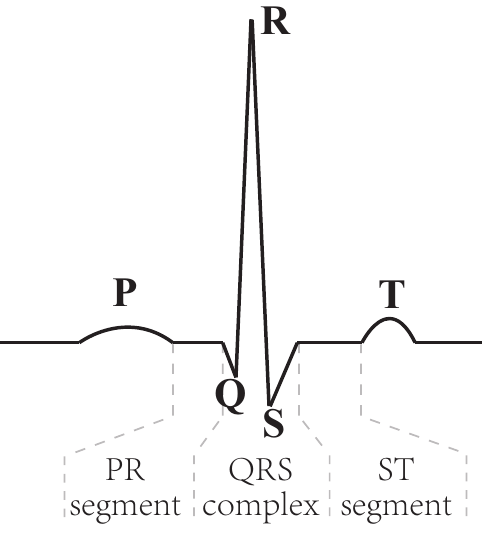}
\caption{A typical ECG signal.}
\label{6}
\end{center}
\end{figure}

Despite the improved accuracy of DNNs, their black-box nature makes them unsuitable for clinical practice --- the inner reasoning and decision-making processes of DNNs remain opaque~\cite{15}. To address this challenge, the so-called attention maps were introduced to explain deep learning inference for detecting acute ICH~\cite{7}. Likewise, attention maps have been adopted to explain the reasoning behind deep learning-based decision-making in ECG analysis~\cite{27}. Although attention maps can highlight the focus areas supporting a decision, the explanation is only superficial because the inner reasoning steps are still inaccessible for interpretation.

Instead of attempting to disentangle approximate explanations from a neural network~\cite{Rudin2019}, we here intend to solve the transparency problem in ECG signal classification by adopting a recent machine learning scheme that is inherently interpretable and transparent, namely, the Tsetlin Machine~(TM)~\cite{16}. TM is a novel machine learning mechanism in which groups of Tsetlin Automata~(TAs)~\cite{18} operate on binary data using propositional logic. The main advantage of TMs for building transparent ECG signal classification is their ability to lay open the reasoning behind the decision-making process. Simultaneously, they are in an increasing number of cases achieving the same or better accuracy than the most recent attention-based neural networks. Enhanced TM architectures~\cite{dropclause,abeyrathna2020massively} have already been successfully utilized in various applications such as aspect-based sentiment analysis~\cite{rohan2021AAAI}, text classification~\cite{Rohanblackbox}, robust interpretation~\cite{yadav2022robustness}, and contextual bandit problems~\cite{RaihanNIPS22}. 

In this paper, we develop an explainable architecture for accurate arrhythmia detection. We design a dedicated TM structure for detecting arrhythmia from long-term ECG signals, and compare the performance with convolutional neural networks on the MIT–BIH Arrhythmia Database. We further propose how to analyse and visualise the decision-making process of the approach, which help us understand the medical basis upon which the architecture makes decisions. To the best of our knowledge, this is the first ML approach that offers both accuracy and interpretability in ECG signal analysis. 

The remainder of this paper is organized as follows. In Section~\ref{Method}, the data set and transparent pre-processing procedure are introduced. Then, in Section~\ref{AI}, we provide the details of our TM-based PVC classification scheme and how we are able to make the ECG analysis interpretable. The numerical results are presented and analyzed in Section~\ref{results}, including a comparison between attention maps and our explanatory TM diagrams, related to medical understanding. We conclude our work in Section~\ref{conclusion}, summarizing our main findings.

\section{Data Set and Pre-processing}\label{Method}

\subsection{The MIT–BIH Arrhythmia Database}
In this study, we use the MIT-BIH arrhythmia database for model training and testing~\cite{28,19}. This database contains long-term (48 half-hour) excerpts of two-channel ambulatory ECG recordings, acquired from 47 subjects (25 men and 22 women). The lead II ECG recordings that we use were sampled at 360 Hz. Eleven of the recordings did not pass the set quality criteria, and were hence excluded from the study. The remaining 36 recordings contain a total of 44180 ECG beats to be classified into five types (Normal, Supraventricular, Ventricular, Fusion and Unknown), according to the recommendations from the Association for the Advancement of Medical Instrumentation (AAMI). In order to distinguish and identify PVC, we further treat the remaining four beat types as Non-PVC beats (38161 beats). To address the multiple forms of PVC~\cite{21}, we divide the PVC category into PVC$_R$ (3965 beats) and PVC$_L$ (2054 beats). Each respectively represents PVC that is caused by irregular contractions starting in the right and the left ventricle~\cite{3}. 

\subsection{Data Pre-processing}
Starting from the MIT–BIH Arrhythmia Database we pre-processed the data according to the following processing steps.

\begin{figure}[t]
\begin{center}
\includegraphics[scale= 1]{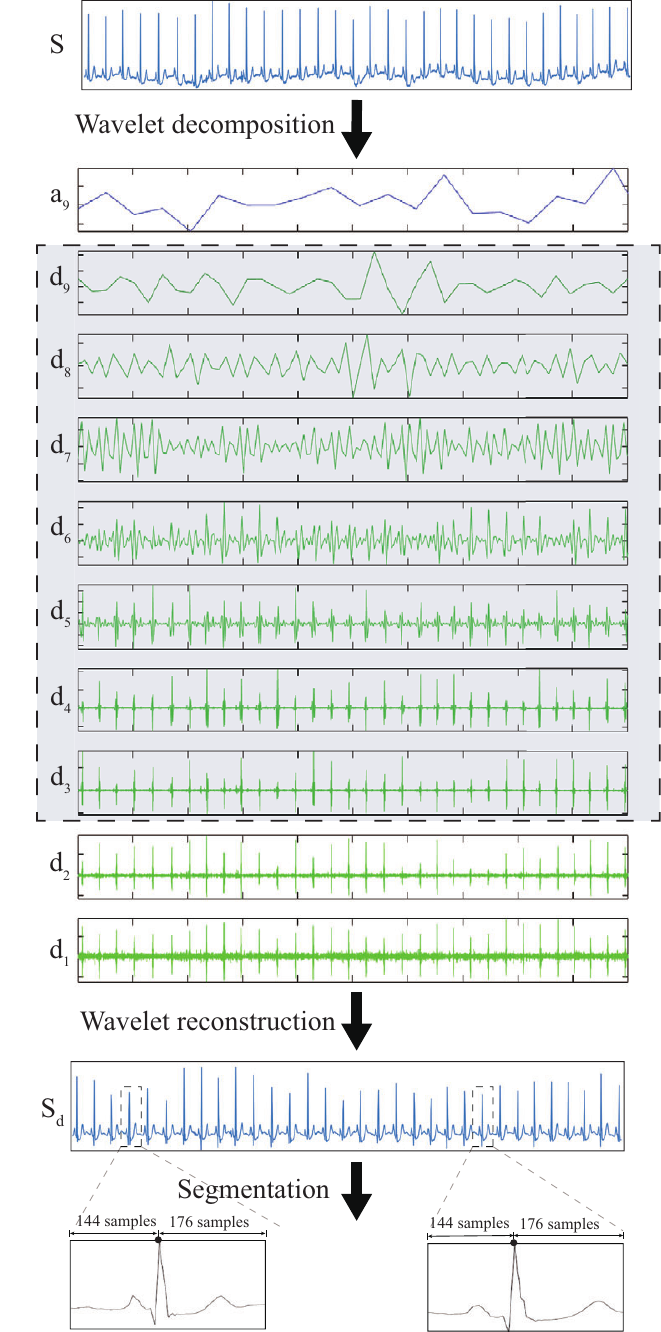}
\caption{The flowchart of data pre-processing.}
\label{1}
\end{center}
\end{figure}

Firstly, the ``Bior 2.6''~\cite{23} wavelet was used to eliminate possible power line interference and baseline wanderings caused by respiration or patient movements. As demonstrated in Figure~\ref{1}, wavelet transform decomposition and reconstruction compensate for the baseline drift of the ECG data. Further, ECG power frequency noise is suppressed, while retaining the change trend of the ECG waveform. As detailed in the figure, the ECG signal was decomposed into nine layers using the ``Bior 2.6'' wavelet, and the resulting high- and low frequency coefficients are extracted. The high-frequency coefficients of the 1st and 2nd layers (d$_1$ and d$_2$ in Figure~\ref{1}) and the low-frequency coefficients of the 9th layer (a$_9$ in Figure~\ref{1}) were then set to $0$ for denoising.

\begin{figure*}[!htb]
\begin{center}
\includegraphics[scale= 0.25]{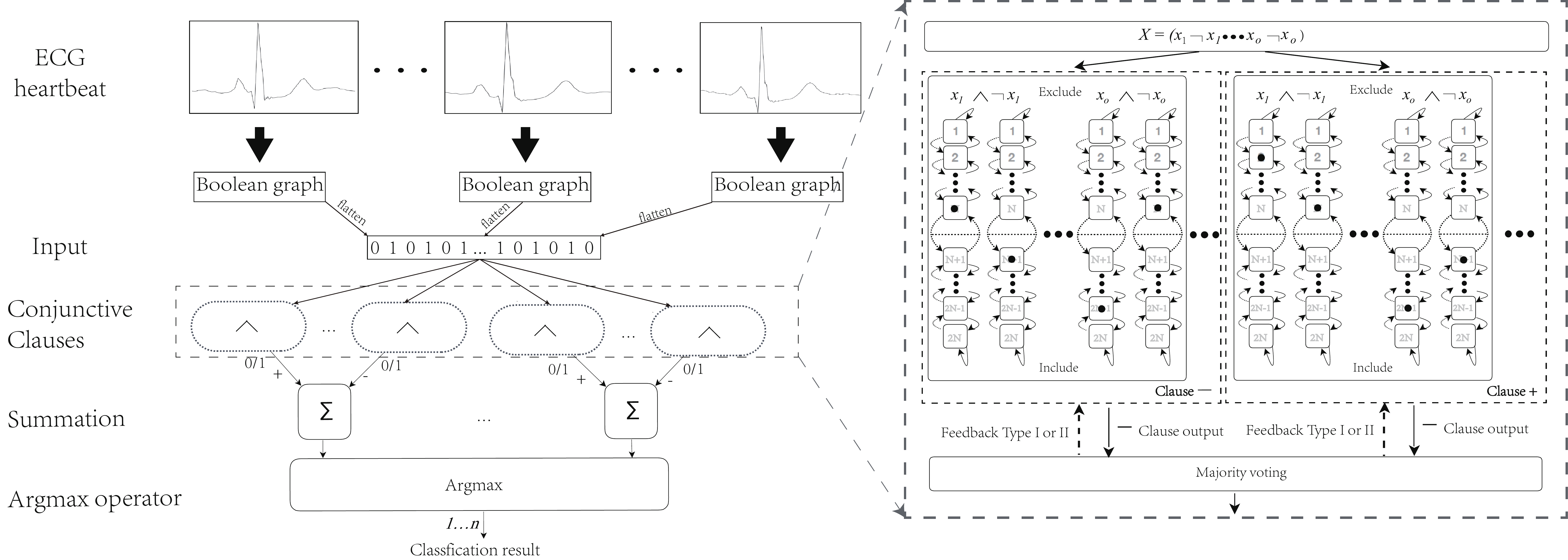}
\caption{Interpretable Tsetlin machine PVC classification architecture.}
\label{2}
\end{center}
\end{figure*}

Secondly, after denoising, in order to divide the ECG signal into single beats to be classified as Non-PVC, PVC$_R$ and PVC$_L$, we segment the signal according to the MIT-BIH reference points (R-peak in the QRS complex of ECG). To ensure that each heartbeat segment contains a complete heartbeat, we select a total of 320 samples (144 samples before and 176 samples after the reference point) as a single heartbeat. Moreover we align the apexes of the QRS complex at the same position on the heartbeat. The segmentation part of the pre-processing is illustrated in the lower part of Figure~\ref{1}.

Finally, each segmented ECG heartbeat was converted into a Boolean-valued two dimensional $100 \times 320$ matrix, where the column index represents time and the row index represents amplitude.  This two dimensional (2-D) matrix is then converted to a one dimensional (1-D) matrix ($1\times 32,000$) by cascading the pixels line by line from top to bottom, which then becomes the input to the TM-based classification model.

\section{Tsetlin Machine PVC Classification Architecture}\label{AI}

The TM is a promising ML algorithm based on propositional logic~\cite{16}, detailed in~\cite{16,17,jiao2021convergenceXOR,jiao2021convergenceAND}. We here briefly go through the basics of TMs. 

%A TM takes a vector $X=[x_1,\ldots,x_o]$ of Boolean features as input, to be classified into one of two classes, $y=0$ or $y=1$. 
Figure~\ref{2} shows the TM structure we propose for PVC classification. As seen, each TM takes a Boolean (propositional) vector $X = (x_1, x_2, \ldots, x_o), x_k \in \{0, 1\}$, $k\in \{1, \ldots , o\}$ as input. In this particular case, the length of the vector is $32,000$. 
From the input vector, we obtain $2o$, i.e., $64,000$ literals $L = (l_1, l_2, \ldots, l_{2o})$. The literals consist of the inputs $x_k$ and their negated counterparts $\overline{x}_k = \lnot x_k = 1-x_k$, i.e., $L = (x_1,  \lnot x_1, \ldots, x_o, \lnot x_o).$

If there are $q$ classes and $n$ sub-patterns in each class, a TM pattern is formulated using \(q \times n\) conjunctive clauses. For any class, we have $n$ clauses\footnote{Note that it is not necessary to have equal number of clauses for each class. The actually number is a configurable hyper-parameter.} indexed by $j$, \(1 \leq j \leq n\):
\begin{equation}
\textstyle
%C_j (X)=\bigwedge_{l_k \in L_j} l_k = \prod_{l_k \in L_j} l_k.
C_j (X)=\bigwedge_{l_k \in L_j} l_k.
\end{equation}
\noindent Here, $L_j$ is a subset of the literals $L$, $L_j \subseteq L$. For example, the clause $C_j(X) = \neg x_1 \land x_2 = (1-x_1) x_2$ consists of the literals $L_j = \{\neg x_1, x_2\}$. The clause outputs $1$ if $x_1 =0$ and $x_2 = 1$, and $0$ otherwise.

A TM forms sub-patterns within a class using $n$ conjunctive clauses $C_j$ (Figure~\ref{2} – Conjunctive Clauses). How the sub-patterns relate to the classes is captured by assigning polarities to the clauses. Positive polarity is assigned to one half of the clauses, denoted by $C^+_j$. These are to capture sub-patterns belonging to the target class ($y=1$). Negative polarity is assigned to the other half, denoted by $C^-_j$. Negative polarity clauses are to capture the sub-patterns for the non-target class  ($y=0$). In effect, the positive polarity clauses vote for classifying the input as the target class, while negative polarity clauses vote against. 

The final classification decision is carried out by summing up the clause outputs (Figure~\ref{2} - Summation). That is, the negative outputs are substracted from the positive outputs. Employing a single TM, the sum is then thresholded using the unit step function $u, u(v) = 1$ \textbf{if} $v \ge 0 \ $\textbf{else} $0$, as shown in Eq.~(\ref{eqn:output}):

     \begin{equation}
      \hat{y} =  u \left (  \sum_{j=1}^{n/2}C^+_j(X) - \sum_{j=1}^{n/2}C^-_j(X) \right).\label{eqn:output} 
    \end{equation}
    
\begin{figure*}[!hbt]
\begin{center}
\includegraphics[scale= 0.45]{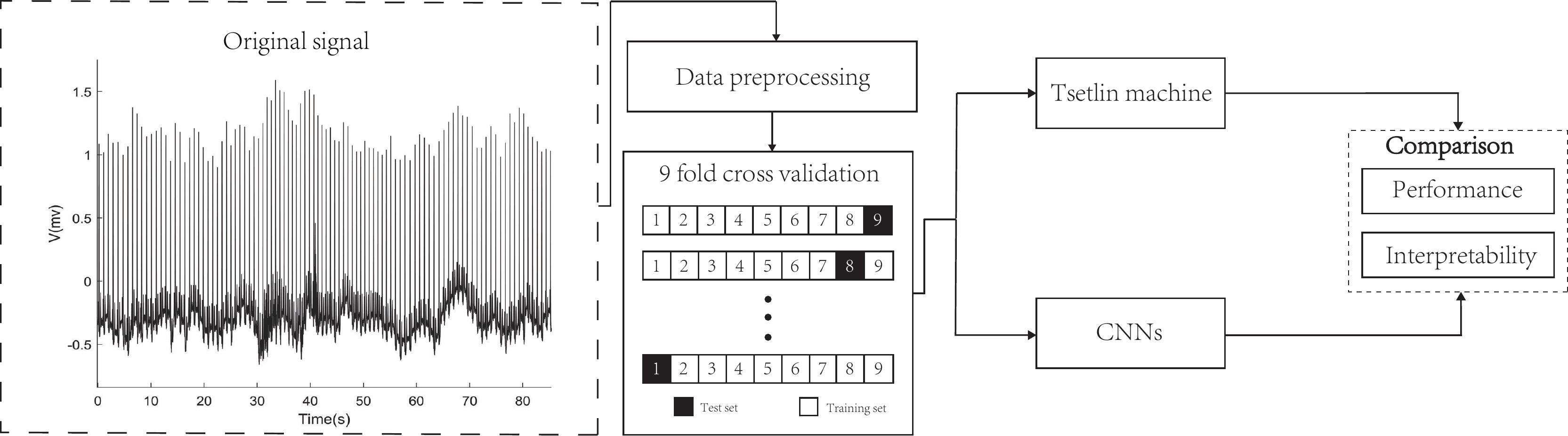}
\caption{Overall flow chart of the experiment.}
\label{flowchart}
\end{center}
\end{figure*}

For training, the key learning component is the TA~\cite{18}. More specifically, a clause $C_j(X)$ is composed by a team of two-action TAs, each TA deciding to \emph{Include} or \emph{Exclude} a specific literal $l_k$ in the clause. Each TA makes decision based on the feedback it receives in the form of Reward, Inaction, and Penalty. The reinforcement depends on six factors: (1) target output ($y = 0$ or $y = 1$), (2) clause polarity, (3) clause output ($C_j$ = 0 or 1), (4) literal value ($x=1$ or 0; or $¬x=1$ or 0), (5) the current vote sum, and (6) the current state of the TA. The TM learning process carefully guides the TAs to converge towards optimal decisions. The details of the training process can be found in ~\cite{16}.

%The TM learning process carefully guides the TAs to converge towards optimal decisions.To this end,the TM organizes the feedback it gives to the TAs into two feedback types. Type I feedback is designed to produce frequent patterns, combat false negatives, and make clauses evaluate to 1. Type I feedback is given to positive polarity clauses when y = 1 and to negative polarity clauses when y = 0. Type II feedback, on the other hand, increases the discriminating power of the patterns, suppresses false positives, and makes clauses evaluate to 0. Type II feedback is given to positive polarity clauses when y = 0 and to negative polarity clauses when y = 1. The feedback is further regulated by the sum of votes $v$ for each output class. That is, the voting sum is compared against a voting margin $T$,which is employed to guide distinct clauses to learn different sub-patterns. The details of the learning process can be found in ~\cite{16}.

Now let us revisit the ECG images. Here, each Boolean-valued pixel is represented by two literals, original form and negated, yielding two TAs. After training, the TA-pair concludes whether to include the pixel in its original form or in its negated form, or to exclude the pixel from the corresponding clause. In this way, each clause can represent a certain sub-pattern of the class. The sub-pattern is in conjunction form, so that a complex pattern can be composed from 0-valued and 1-valued pixels. Thereafter, the TM clauses can make a joint classification decision based on the voting for and against the classes. 

\section{Experiment Results}\label{results}

We now evaluate our TM-based architecture, comparing it against CNNs.  The overall experiment setup is shown in Figure~\ref{flowchart}, and detailed in the following.

\subsection{Experiment Setup}

We train and evaluate the TM with the filtered and segmented MIT-BIH data using nine-fold cross-validation. As a baseline, we use a deep 1-D CNN trained under the same experiment conditions as in~\cite{25}. The deep learning model is optimized to provide high recognition performance on the ECG signals using standard CNN layers~\cite{25}. The 1-D CNN model consists of ten layers, with two 1-D Convolution layers, one Maximum pooling layer, one Dropout layer, two 1-D Convolution layers, one Maximum pooling layer, one Flattening layer, one Dense layer, and finally, a SoftMax layer. The specific parameters, such as the number of convolution kernels, strides and activation function, can be found in~\cite{25}. 
 {\color{black}Additionally, we also adopted a widely-used 2-D CNN model (VGG16)~\cite{30} for performance and interpretability comparision. }

All experiments are conducted on a Linux Server (Ubuntu 20.04.4) with NVIDIA GeForce GTX 3090 (24GB). We ran our TM for 150 epochs with a hyperparameter configuration of $5000$ clauses, margin $T=5000$, and specificity $s= 1.5$ (see \cite{16} for a detailed explanation of the hyperparameters). The PyTorch library realizes the deep learning algorithms~\cite{26}.  We use the Adam optimizer to train the CNN model, with categorical cross entropy loss, a learning rate of 10$^{-3}$, and decay 10$^{-4}$. The batch size is $64$ and each training phase encompasses $100$ epochs.

\subsection{Classification Results}

In order to robustly estimate the generalization capability of the model, the segmented ECG heartbeats from all the 36 subjects were divided into 9 folds randomly (4 subjects per fold). All estimates were then found using 9-fold cross-validation. 

%For the evaluation of TM and CNN, we assumed the class containing Non-PVC, PVC$_L$ and PVC$_R$, and therefore the accuracy,  Precision, and recall are computed with reference to this.
The classification performance is evaluated using \emph{accuracy}, \emph{precision}, and \emph{recall}, defined below:
    \begin{eqnarray}
      \mathit{Accuracy}&=&\dfrac{TP + TN }{TP+TN+FP+FN},\\
      \mathit{Precesion}&=&\dfrac{TP }{TP+FP},\\
      \mathit{Recall}&=&\dfrac{TP }{TP+FN}.
    \end{eqnarray}
Here, \textit{True Positive} (TP) is the number of samples that were originally positive for the category and were correctly classified. \textit{True Negative} (TN) is the number of samples that were originally negative for the category and were classified as negative. \emph{False Positive} (FP) is the number of samples that were originally negative for the category but were classified as positive. And, \emph{False Negative} (FN) is the number of samples that were originally positive for the category, but were classified as negative.

\begin{figure*}[!hbt]
\begin{center}
\includegraphics[scale=0.55]{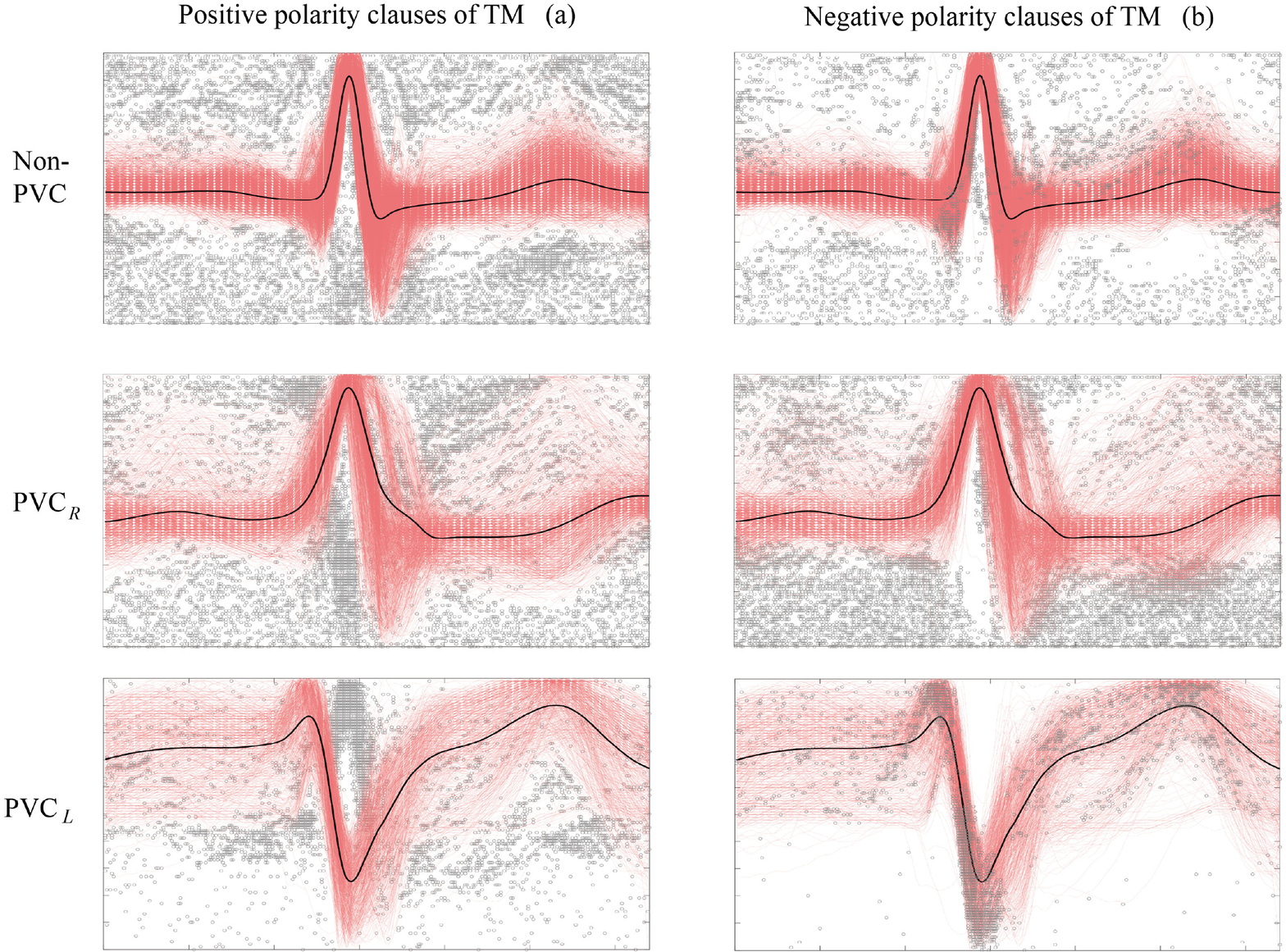}
\caption{Interpretability maps of TM produced by plotting the role each pixel plays in each clause. The positive polarity clauses are to the left (a) and the negative polarity clauses are to the right (b). The thin red lines show the samples of the ECG heartbeat waveforms in each category. The thick
black line is the average of these. The symbol ``o" signify that the status of corresponding pixel is ``0", while a blank pixel states that the pixel is ``Excluded" in the displayed clauses.}
\label{pixel}
\end{center}
\end{figure*}

\begin{table}
    \caption{Classification performance of TM and CNNs.}
    \centering
    \begin{tabular}{ccccc}
        \hline
        Model  & Category & Precision & Recall &Accuracy\\
        \hline
\multirow{3}*{1-D CNN}&  Non-PVC & 97.8\% & 95.0\% & \multirow{3}*{93.8\%}\\
                     & PVC$_R$ & 73.0\%   & 79.5\% \\
                     & PVC$_L$ & 96.2\%   & 96.1\% \\
        \hline
\multirow{3}*{VGG16}&  Non-PVC & 96.8\% & 96.8\% & \multirow{3}*{94.2\%}\\
                     & PVC$_R$ & 70.7\%   & 68.7\% \\
                     & PVC$_L$ & 90.3\%   & 92.6\% \\
        \hline
 \multirow{3}*{TM}&  Non-PVC & 97.4\% &  95.9\% & \multirow{3}*{94.2\%}\\
                  & PVC$_R$ & 72.1\%   & 74.3\% \\
                  & PVC$_L$ & 93.0\%   & 97.7\% \\
                   \hline
    \end{tabular}
    \label{tab:plain}
\end{table}

Table~\ref{tab:plain} reports the outcome of the TM and the neural networks for the targeted datasets. Compared with the 1-D CNN, the TM obtained 94.2\% in accuracy, which is better than the 1-D CNN accuracy of 93.8\%. TM further gave 95.9\% and 97.7\% in sensitivity towards Non-PVC and PVC$_L$ respectively, which outperformed the 1-D CNN. For the other metrics, the performance of TM is close to those of the 1-D CNN. Compared with the VGG16, TM outperforms it in most of metrics.  In conclusion, the numerical results indicate that the TM performs comparably to the deep 1-D CNN and the VGG16, at a level that opens up for application in long-term ECG monitoring for PVC identification.

\begin{figure*}[!ht]
\begin{center}
\includegraphics[scale= 0.9]{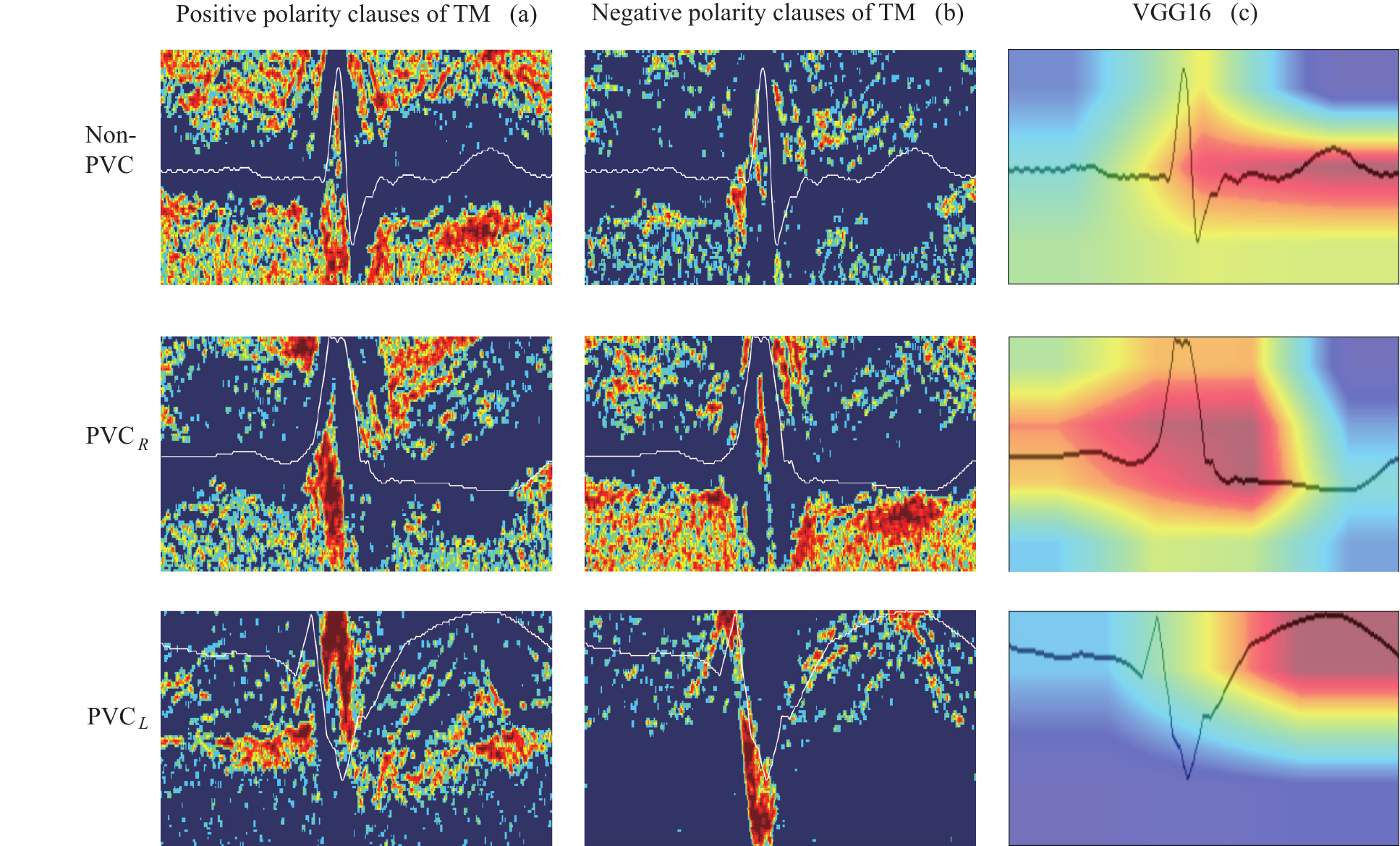}
\caption{ Heatmap Comparison of TM and VGG16.}
\label{heatmap}
\end{center}
\end{figure*}

\subsection{Interpretability Analysis}

In this application, the TM clauses form self-contained patterns by joining pixels into multi-pixel structures. That is, the image pixels are inputted directly to the clauses, which are propositional AND-rules composing pixel patterns.  With the patterns expressed directly in multi-pixel structures, they become directly accessible for human interpretation. Additionally, the AND-form is particularly suitable for human comprehension~\cite{24}.

Each ECG signal category gets its own pool of clauses. One half of these describes ECG curve patterns that characterize the category. These are the positive polarity clauses and are used to determine whether the ECG signal belongs to the category. The other half (the negative polarity clauses) describe curve patterns that are uncharacteristic for the category. These are used to judge whether the signal belongs to some other category.

Implementation-wise, each clause operates on a binary ECG heartbeat matrix. A clause then specify a role for each pixel, which is ``1", ``0'', or ``*''. For positive polarity clauses,  ``1" means that the corresponding pixel must be 1 for the heartbeat to belong to the clause's class (the pixel is included in original form). Oppositely, ``0" states that the corresponding pixel must be 0 to fit into the class (the pixel is included in negated form). The ``*"-role, on the other hand, means that the clause does not consider this pixel (the pixel is excluded from the clause). In other words, the pixel value does not influence whether the clause matches the ECG signal. Following the same procedure, the negative polarity clauses characterize the other classes, again using the pixel roles ``1"，``0"，``*".

%when clauses capture patterns from a segmented ECG heartbeat binary matrix, each pixel of the matrix is defined as three states, ``1", ``0" and ``*". For even clauses ``1" represents that the corresponding pixel must be 1 when the heartbeat was classified as its category, while, ``0" represents that the corresponding pixel must be 0. And, ``*" means that the pixel could be 0 or 1 either. Meanwhile, for odd clauses, ``1"，``0"，``*" represent the corresponding pixel points with the same meaning as in even clauses, but for other categories. 

 To visualize the interpretability of TMs, we plot the pixel patterns of all the trained clauses in Figure~\ref{pixel}. There are six plots because we have three categories and each category gets two sets of clauses (positive and negative polarities). In these plots, we use the symbol ``o" to represent pixel value ``0" (the negated form of a pixel). For excluded pixels, instead of using ``*", we leave the pixel blank to avoid cluttering the plot. Notice that in our case, the clauses only include pixels in negated form. Hence, the plots contain ``o"s, exclusively. Indeed, with sparse data and small specificity $s$, negated features are more typical than plain features, such as for natural language processing~\cite{rohan2021AAAI}.
 
Figure~\ref{pixel} captures the following salient properties:
\begin{itemize}
\item For Non-PVC waveforms, the symbol ``o" is distributed evenly around the ECG waveforms in the map for positive polarity clauses (a). Accordingly, Non-PVC ECG waveforms should not appear in these areas. In the map depicting negative polarity clauses (b), we observe groups of ``o" distributed below the R-wave. These show that the waveform of PVC$_R$ and PVC$_L$ (i.e., categories that are not Non-PVC) cannot appear in these areas for the clauses to match. Accordingly, the Non-PVC waveforms have a narrow QRS complex compared to PVC, which is consistent with the clinical standard~\cite{3}.
\item For the PVC$_R$ waveforms, we observe that the positive polarity clauses (a) form a high density distribution of ``o"s under the R-wave. This distribution means that the PVC$_R$ waveform cannot cross these areas if the clauses are to match. In addition, it is clear that the PVC$_R$ has a wide QRS complex, especially a wide R-wave. Indeed, this is one of the evidences clinicians use to diagnose PVC~\cite{3}.
\item For PVC$_L$, the clinical evidence is a wide and negative QRS complex followed by a positive ST-T segment. As seen in the map for positive polarity clauses of PVC$_L$, the groups of ``o"s are distributed above and inside the negative R-wave with relatively high density, showing that PVC$_L$ has wider negative R-wave than Non-PVC and a significant negative QRS complex compare with PVC$_R$. In the map for the negative polarity clauses, we can see that a concentration of ``o"s appears on the falling edge of the negative R-wave and the peak of the positive S-wave and T-wave. This pattern means that Non-PVC and PVC$_R$ waveforms do not appear in these areas. In conclusion, the pattern indicates that PVC$_L$ waveforms indeed have negative QRS complex and positive T-wave~\cite{3}.
\end{itemize}

From the above analysis, we conclude that our TM architecture can provide pixel-level interpretability that is compatible with medical knowledge. The positive and negative polarity clauses play distinct roles in differentiating between the different types of ECG heartbeat. That is, each of the six clause types highlight different kinds of medical evidence. Most importantly, the visualized decision-making processes corresponds closely with that of a clinician.

\subsection{TM vs CNN Interpretability}
To cast further light on TM interpretability, we now use heatmaps to contrast TM clauses against CNN attention. A heatmap visualizes the importance of a phenomenon in 2-D using color. To compare the difference between the basis for TM decisions and the basis for CNN decisions, we employ VGG16 and plot its heatmap using Gradient-weighted Class Activation Mapping (Grad-CAM)~\cite{31}. The heatmap was produced according to the output layer of VGG16, showing the importance of each pixel with regards to deciding the output.

%We have noticed that in recent years, when the convolutional neural network is used for image classification, it can draw the corresponding heatmap by taking out the internal weight value to show the basis of CNN classification. Therefore, we hand over one of the folds of this data to the VGG16 network for classification, and draw heatmaps through the weight of the feature layer (VGG16 has a classification accuracy of 96\% for this fold)

To make the interpretability comparison compatible, we visualize the interpretability maps from Figure~\ref{pixel} as heatmaps. In these heatmaps, the color measures the density of clauses that include each pixel in negated form (pixel value ``0") in logarithmic scale. We produce the map using Eq.~(\ref{logplot}), scanning the whole interpretability map with a $3\times3$ window and step size 1. We refer to the density of ``0"s in the window as local logarithmic density (LLD):
    \begin{equation}
      LLD = 10\log_{10}\dfrac{\mathit{Number\;of}\;``0"s}{9}. \label{logplot}
    \end{equation}
We then plot the obtained densities in color scale as shown in Figure~\ref{heatmap}(a) and Figure~\ref{heatmap}(b). The attention heatmaps for VGG16 are in  Figure~\ref{heatmap}(c).
We finally select one typical ECG heartbeat waveform from each category, adding them to the heatmaps. 

From the heatmaps, we observe the following:
\begin{itemize}
\item For Non-PVC, VGG16 focuses on the ST-T segment to classify a normal heartbeat. In contrast, the TM focuses on the width of the R-wave.
\item For PVC$_R$, VGG16 focuses on almost the whole of the QRS complex area. The TM, on the other hand, pays more attention to the width of the R-wave (heatmap for positive polarity clauses). We can also observe that the TM focuses on the area blow the T-wave (in the heatmap of the negative polarity clauses).
\item For PVC$_L$, it is clear that VGG16 focuses on the T-wave to make its decision, while the TM makes its decision based on to the wide and negative QRS complex plus the positive T-wave.
\end{itemize}
According to the above observations, the heatmaps of VGG16 tell us only where the attention is, but not what the attention means (why is the attention of medical importance). In contrast, the TM tells us both where to focus (the distribution of the included pixels) and why because of the six types of clauses. Each type corresponds to medical knowledge that explains the focus. That is, when the TM recognize Non-PVC, PVC$_R$ and PVC$_L$, the decision is founded on the narrow QRS complex, wide R-wave and the negative QRS complex, respectively, which indicate that the TM interpretation is more meaningful than CNN-based VGG16 attention. Indeed, TM interpretability seems to be consistent with the clinical standard.

\section{Conclusion}\label{conclusion}

This paper proposed how the TM can be used to accurately recognize ECG signal classes in an interpretable manner. Introducing TM interpretability maps and heatmaps, we were able to illustrate the decision-making process of TMs. One key finding was that the TM described the medical patterns through two distinct lenses: positive and negative polarity clauses. The positive polarity clauses characterized what the target class looked like, while the negative polarity clauses described what the target class should not look like. Together, the two kinds of clauses differentiated their target class from the other classes. From the experiments, we found that the TM was capable of producing human-interpretable rules, which were consistent with the clinical standard. In addition, the performance of TM was comparable with deep CNN-based models. Given the competitive accuracy, transparency, and correspondence with medical knowledge, we believe TMs can be utilized to build trustworthy AI systems for real-life clinical use. %Future works include (1) extending its usage to more physiological signals such as electromyogram (EMG), electroencephalogram (EEG), photoplethysmogram (PPG) and phonocardiogram (PCG) and (2) optimizing its computational efficiency.

%\section*{Acknowledgments}
%This work was supported in part by the key research and development program of Sichuan province under Grant 2022YFG0045; in part by the project Spacetime Vision: Towards Unsupervised Learning in the 4D World financed by the EEA and Norway Grants 2014–2021 under the Grant No. EEA-RO-NO-2018–04; in part by Fundamental Research Funds for the Central Universities under Grant 2022SCU12008 and in part by National Natural Science Foundation of China under Grant 62066042.

%% The file named.bst is a bibliography style file for BibTeX 0.99c
\bibliographystyle{named}
\bibliography{New}

\end{CJK}
\end{document}